\documentclass{article}
\usepackage[margin=1in]{geometry} % Adjust margin size as needed
\usepackage{graphicx} % Required for inserting images
\usepackage{hyperref}
\usepackage[numbers]{natbib}
\usepackage{amsmath}
\usepackage{xcolor}
\usepackage{booktabs} 
\usepackage{authblk} % For handling multiple authors and affiliations
\usepackage{placeins}
\usepackage{pdfpages}
\newcommand{\fig}[1]{\textcolor[rgb]{0,0.5,0}{#1}}

\newcommand{\tdeux}{T\textsubscript{2}}

\begin{document}

% \begin{center}
%     \Huge \textbf{Final version available in Magnetic Resonance in Medicine DOI: 10.1002/mrm.30434 }
% \end{center}

\begin{center}
    \Huge \textbf{Submitted to Magnetic Resonance in Medicine}
\end{center}

\vspace{1em}

\title{\textbf{MR-WAVES: MR Water-diffusion And Vascular Effects Simulations\protect}}

\date{March 2025}

\author[1]{Thomas Coudert}
\author[1]{Maitê Silva Martins Marçal}
\author[2]{Aurélien Delphin}
\author[1]{Antoine Barrier}
\author[1]{Lila Cunge}
\author[1,3]{Loïc Legris}
\author[1]{Jan M. Warnking}
\author[1]{Benjamin Lemasson}
\author[1]{Emmanuel L. Barbier}
\author[1]{Thomas Christen}

\affil[1]{Univ. Grenoble Alpes, INSERM, U1216, Grenoble Institute Neurosciences, GIN, Grenoble, France}
\affil[2]{Univ. Grenoble Alpes, INSERM, US17, CNRS, UAR3552, CHU Grenoble Alpes, IRMaGe, Grenoble, France}
\affil[3]{Univ. Grenoble Alpes, Stroke Unit, Department of Neurology, CHU Grenoble Alpes, Grenoble, France}

\maketitle

\begin{abstract}
Accurate MR signal simulation, including microvascular structures and water diffusion, is crucial for MRI techniques like fMRI BOLD modeling and MR vascular Fingerprinting (MRF), which use susceptibility effects on MR signals for tissue characterization. However, integrating microvascular features and diffusion remains computationally challenging, limiting the accuracy of the estimates.  Using advanced modeling and deep neural networks, we propose a novel simulation tool that efficiently accounts for susceptibility and diffusion effects. We used dimension reduction of magnetic field inhomogeneity matrices combined with deep learning method to accelerate the simulations while maintaining their accuracy. We validated our results through an in silico study against a reference method and in vivo MRF experiments. This approach accelerates MR signal generation by a factor of almost 13,000 compared to previously used simulation methods while preserving accuracy. The MR-WAVES method allows fast generation of MR signals accounting for microvascular structures and water-diffusion contribution.
\end{abstract}

\section{Introduction}\label{intro}

The diffusion of water molecules in biological tissues is a fundamental process influencing MRI signals. In homogeneous space, diffusion follows an isotropic Brownian motion described by Fick’s law, with a characteristic diffusion coefficient specific to the medium\cite{albert1905a, crank_mathematics_1979}. However, in biological tissues, cell membranes, microvascular structures, and other barriers lead to restricted and anisotropic diffusion, altering the signal properties in diffusion-sensitive MRI sequences\cite{le_bihan_mr_1986}.

In diffusion-weighted MRI, the effect of diffusion is quantified by signal attenuation in the presence of diffusion-sensitizing gradients, allowing the measurement of apparent diffusion coefficients (ADC) that reflect tissue microstructure. Similarly, in susceptibility-based MRI techniques such as blood oxygenation level-dependent (BOLD) functional MRI\cite{ogawa_brain_1990} and dynamic susceptibility contrast (DSC)\cite{ostergaard_high_1996} imaging, diffusion plays a key role in modulating the phase and magnitude of MR signals in the presence of microscopic magnetic field inhomogeneities. The interaction between water diffusion and local field variations around microvascular structures influences contrast mechanisms, making diffusion an essential factor in accurately modeling and interpreting MR signals in advanced imaging methods.

Accurately simulating MR signals in the presence of microvascular structures and water-diffusion effects plays an important role in the development of new imaging biomarkers helping to understand brain physiology and improve diagnostic precision. This is the case for MR vascular Fingerprinting (MRvF \cite{christen_mr_2014}) where microvascular simulations are directly compared to in vivo MR signal time courses. However, the computational burden of such simulations remains a critical bottleneck for the generation of large-scale dictionaries. Initial MRvF studies \cite{lemasson_mr_2016, delphin_enhancing_2024} utilize a GESFIDSE sequence (Gradient-Echo Sampling of Free Induction Decay and Spin Echo)\cite{ma_method_1996} pre- and post-USPIO injection, comparing concatenated acquired signals with a "dictionary" of simulated signals. This approach enables the quantification of Cerebral Blood Volume (CBV), mean vessel radius (R), relaxation time T2, and oxygenation (SO$_2$). Dictionary generation relies on a previously published tool\cite{pannetier_simulation_2013} that accounts for susceptibility \cite{salomir_fast_2003} and diffusion. Unlike previous diffusion models for which effects were simulated using random walk (Monte Carlo) methods\cite{ogawa_functional_1993, weisskoff_microscopic_1994, yablonskiy_theory_1994, boxerman_intravascular_1995}, here a deterministic convolution process is performed using a Gaussian model of motion, where the random displacements of water molecules due to diffusion are modeled as a Gaussian distribution of displacements over time \cite{bandettini_effects_1995}. A Gaussian kernel representing the probability density function of displacements is convolved with the MR signal at each small time step (<0.5 ms) to capture this process. This convolution allows for the precise accumulation of phase shifts induced by diffusion, ensuring the accurate evolution of the MR signal. The small time step is critical to capture the cumulative effect of microscopic diffusion events accurately, but it also significantly increases the computational burden. Consequently, this approach requires intensive computations, leading to long simulation times (often days for thousands of signals), which constrained prior studies to small-scale dictionaries (less than 30,000 signals). Given the vast diversity of microvascular structures in the human brain and the critical role of water-diffusion effects in ensuring simulation realism, there is a pressing need for more efficient simulation tools.

In this work, we introduce a fast method for Water-diffusion and Vascular Effects Simulations (MR-WAVES). Our contribution is two-fold: First, as already introduced in a previous study \cite{coudert_relaxometry_2025}, we accelerate intra-voxel microvascular susceptibility calculations by moving from a three-dimensional (3D) to a one-dimensional (1D) representation. This time, we also investigate the accuracy of the 1D histogram approximation of 3D magnetic field inhomogeneity matrices, depending on the sampling parameters such as the bins and the edges of the distribution. Secondly, we introduce a direct approach to diffusion using a Recurrent Neural Network (RNN), which transforms a zero-diffusion simulated signal based on specified microvascular properties and a fixed ADC value. The combined approach removes the need for constant time steps and, similar to Bloch simulations, computes the magnetization only when events (RF, gradient...) occur.

To validate our method, we compare MR-WAVES against the described brute-force method, highlighting the accuracy and computational efficiency trade-offs. Finally, we will demonstrate the applicability of our method on MR vascular Fingerprinting pre-clinical data, showcasing its potential to enable larger, more comprehensive analyses of vascular properties in the brain.

%%%%%%%%%%%%%%%%%%%%%%%%%%%%%%%%%%%%%%%%%%%%%%%%%%%%%%%%%%%%%%%%%%%%%
%%%%%%%%%%%%%%%%%%%%%%%%%%%%%%%%%%%%%%%%%%%%%%%%%%%%%%%%%%%%%%%%%%%%%
%%%%%%%%%%%%%%%%%%%%%%%%%%%%%%%%%%%%%%%%%%%%%%%%%%%%%%%%%%%%%%%%%%%%%

\section{Methods}\label{matmet}

\subsection{Computations of magnetic field inhomogeneities} \label{microvasc}
3D microvasculatures were segmented from microscopy-imaged batches of a healthy mouse brain. The microscopy dataset\cite{di_giovanna_whole-brain_2018} of mice brain voxels was processed to obtain a set of MRI-sized microvascular voxels (\textit{i.e.} 248×248×744$\mu m^3$) and segmented using the VesselExpress software\cite{spangenberg_rapid_2023} to derived binary masks representing the vascular network (see \fig{Figure 1 left}). Voxels were then rescaled to 2×2×2$\mu m^3$ resolution and characterized in terms of total CBV and R, using the same software. To achieve a large diversity of voxels, representing healthy and pathological vascular networks, two types of data augmentation were performed as illustrated in \fig{Figure 2}: dilation, using two different structuring elements: a 3x3x3 cube (\fig{Figure 2a middle}) and a sphere with a radius size of 2 (\fig{Figure 2a right}); and merging (\fig{Figure 2b}), where two or more segmented MR-size voxels were superimposed (right panel). From 20,900 voxels, around 3,500 merged voxels were added, and around 20,600 dilated voxels. In the end, this resulted in 45,000 segmented voxels. 

Magnetic susceptibility maps were computed with ($\chi_{USPIO}$=3.5ppm) and without the presence of a USPIO (Ultrasmall superparamagnetic iron oxide) contrast agent (CA) (see \fig{Figure 1}). The SO$_2$ and hematocrit values are fixed for each voxel. From this, 3D magnetic field maps were computed in Python 3.8.10 for each voxel using a Fourier method \cite{salomir_fast_2003}. The simulations were performed for a static magnetic field strength of 4.7T.

% \begin{figure}[!ht]
%     \centering
%     \includegraphics[width=\textwidth]{Figures/chap5/augm_maite.png}
%     \caption[Microvascular voxel data augmentation.]{Microvascular voxel data augmentation. Example of dilation process (left) and merging process (right). Courtesy of Maitê Silva Martins Marçal.}
%     \label{fig:ch5:augm_maite}
% \end{figure}

\subsection{MR Sequence}

MR simulations were performed for a GESFIDSE sequence with 32 gradient echoes, before and after USPIO contrast agent injection. All the following numerical simulations of MR signals are made separately for pre and post-CA signals. For the MRF in vivo study (see \ref{MRF}), pre and post-CA signals were concatenated before reconstruction, resulting in a set of 64-timepoints signals.

\subsection{Reference: 3D simulations}
As a reference in our study, 3D matrices of magnetic field inhomogeneities were computed using a ﬁrst order perturbation approach to Maxwell’s magneto-static equation, combined with the Fourier transformation technique\cite{salomir_fast_2003}. Using the fully 3D matrices of size 248x248x744, the MR simulations were carried out in Matlab R2024a \cite{MATLAB}, using parallel computation of 64 CPU cores and intensive RAM allocations\cite{delphin_enhancing_2024} without and with water-diffusion effect. In the latter case, an apparent diffusion coefficient (ADC) of 850$\mu\text{m}^2/\text{s}$ was chosen, considering it as a mean value in healthy brain tissues \cite{helenius_diffusion-weighted_2002}. Assuming that the motion of water molecules follows a normal distribution in space\cite{bandettini_effects_1995}, diffusion effects were modeled by the convolution of a Gaussian kernel. To ensure the accuracy of the motion simulation, computations were made using a 0.5ms time step. For computational efficiency, signals were firstly computed for a fixed \tdeux{} value of 100ms. Varying \tdeux{} effects were then added \textit{a posteriori} by multiplying each GESFIDSE signal by the appropriate exponential decay functions\cite{delphin_enhancing_2024}. 

In order to maintain reasonable computation time, 20,000 microvascular structure voxels among the 45,000 ones, combined with Sobol\cite{sobol_distribution_1967} distributed values of SO$_2$ and \tdeux{}, were used to generate two 20,000-signal sets, with and without diffusion effect using the brute force approach. We respectively called these methods \textit{Ref}$_{diff}$ and \textit{Ref}$_{nodiff}$ in the following of this article (see \fig{Figure 4}). 

\subsection{MR-WAVES approach}

The source code for our implementation is available \href{https://github.com/ThomasCoudert/MR-WAVES-public}{here}. The proposed MR-WAVES approach, as introduced, combines two modules that are separately detailed in the two following subsections.

\subsubsection{1D simulations without diffusion effect}

Using the Larmor equation, the 3D magnetic fields ($\delta B$) maps derived from microvascular voxel can be seen as 3D frequency inhomogeneities ($\delta f$) maps.
\begin{equation}
    \delta f = \frac{\gamma}{2\pi} \cdot \delta B
\end{equation}
\text{where } \(\gamma\) \text{ is the gyromagnetic ratio (e.g., } \(\gamma / 2\pi = 42.58\) \text{MHz/T for protons).}

To simplify and accelerate the computations, we approximated this 3D frequency matrix using a 1D histogram, neglecting spatial arrangements and assuming that the frequency offsets can be described statistically rather than spatially\cite{coudert_relaxometry_2025}. The 1D histogram was sampled using the \texttt{numpy.histogram} function in Python, providing a discrete approximation of the frequency distribution. From this dimension reduction, complex single-frequency MR signals obtained with Bloch equation simulations\cite{hargreaves_bloch_nodate} performed independently for different isochromats were combined according to the resulting 1D distribution histograms. This allowed for fast computation of MR signals at time steps defined by sequence events while accounting for the contribution of microvascular properties (SO$_2$, CBV, R). We called this approach $MRWAVES_{nodiff}$ because it doesn't include the water-diffusion effect (see \fig{Figure 4}).

Simulations were performed with a mix of Python 3.8.10 and Matlab R2024a, using parallel computation on 64 CPU cores and intensive RAM allocations.

\subsubsection{Fast simulations of the water diffusion effect on MR signals using neural networks}

In the brute force approach, the effect of water diffusion was modeled by the convolution of a Gaussian kernel updating the MR signal evolutions each 0.5ms time step, which is highly time-consuming and limits the number of MR signals that can be generated. Yet, the effects of water diffusion that we have observed on the GESFIDSE signals are mainly a shape modification between the signal without and the signal with diffusion.  

Recurrent Neural Networks (RNNs) are well-suited for sequential data \cite{elman_finding_1990}, therefore MRI time courses\cite{liu_fast_2021, sobczak_predicting_2021, oksuz_magnetic_2018} as they maintain a memory of previous states, allowing them to model temporal dependencies in time-series data. While Long Short-Term Memory (LSTM) networks have already been successfully applied in MR Fingerprinting for signal modeling or parameters estimation\cite{cabini_fast_2024, hoppe_rinq_2019, barrier_marvel_2024}, we opted here to use Gated Recurrent Units (GRU) due to their simpler architecture and faster training times. GRUs retain the key capability of handling long-term dependencies while being computationally more efficient, making them well-suited for our application. We implemented and trained a GRU RNN to model the water diffusion effect on MR signals as a surrogate for time-consuming deterministic methods. To avoid any bias, training and testing of the GRU-RNN were conducted using signals simulated from the brute-force methods \textit{Ref}$_{nodiff}$ and \textit{Ref}$_{diff}$.

\paragraph{Model parameters} The model followed a GRU-RNN architecture (see \fig{Figure 3}) with a single output dimension representing the magnetization vector. The network input was a signal without water diffusion effects, along with a set of (SO$_2$, CBV, R) tissue parameter values. The network processed batches of 5 signals, while the RNN layer contained 32 units. An Adam optimizer, a popular gradient-based optimization algorithm, was used to optimize the model with a Mean Squared Error (MSE) loss. Learning rate adjustments were handled such that the learning rate was reduced when the validation loss plateaued, using a factor of 0.8 and patience of 20 epochs. Additionally, a minimum learning rate of $1e^{-6}$ was enforced to prevent it from diminishing to very small values. Model checkpoints were saved based on the training loss, with the best model retained in an HDF5 file named according to the experiment setup. Early stopping was employed with a patience of 15 epochs, halting the training process if the validation loss did not improve, to prevent overfitting. Together, these callbacks ensured efficient training while avoiding overfitting and dynamically adjusting the learning rate based on performance trends.

\paragraph{Training set} As for the simulation process, the signals were split into pre- and post-part, and the network was trained in two separate models. In the same way, the prediction is made for pre and post-CA parts of the signals that are then concatenated. The training set was made of 90\% of 20,000 signals from \textit{Ref}$_{nodiff}$ and \textit{Ref}$_{diff}$ \textit{i.e.} 18,000 signals simulated using the brute-force method for the GESFIDSE sequence described previously, once without diffusion effect, and then with diffusion effect to serve as ground truth. 

\paragraph{Testing set} The testing of the model was made on the remaining signals not used for training, \textit{i.e.} 2,000 signals with and without diffusion as the ground truth for evaluating the model. Here we call the method \textit{GRU}$_{diff}$ to specify that the vascular simulations are done using the 3D matrices brute-force method but the diffusion is added with the GRU-RNN (see \fig{Figure 4}).

All model computations were made on an NVIDIA GeForce RTX 3060 GPU.

\subsection{In silico validation study}
\paragraph{MR simulations without diffusion}

We first wanted to assess the accuracy of the dimension reduction technique of the 3D frequency inhomogeneity matrices. For this, we adjusted the sampling strategy in the histogram (\textit{i.e.} its number of bins and its edges) to control the granularity of the frequency representation (\textit{i.e.} the sampling step in Hz) and find a trade-off between accuracy and computational efficiency. We made this with and without UPSIO contribution in the susceptibility computation.

The fully resolved 3D frequency matrices served as the reference for the accuracy evaluation. For both pre- and post-CA simulations, we then compared the set of 20,000 MR signals computed from the brute-force method (\textit{Ref}$_{nodiff}$, using the 3D matrices) and sets of 20,000 MR signals computed with the proposed MR-WAVES method (\textit{MRWAVES}$_{nodiff}$) using different 1D histogram samplings of the 3D matrices.

The accuracy was quantified using the signal-wise Normalized Root Mean Squared Error (NRMSE) between the brute-force signals and those obtained with different sampling parameters. 

\paragraph{MR simulations with diffusion}
Then we compared the proposed method MR-WAVES against the brute-force method, incorporating this time the diffusion effects by testing the trained RNN model on the 2,000 signals. To further investigate the results, this testing was made on the 2,000 signals computed from the brute-force methods using full 3D matrices (\textit{Ref}$_{nodiff}$), and with the 1D distribution histograms (\textit{MRWAVES}$_{nodiff}$). We thus obtained 2,000 output signals under a method we called \textit{MRWAVES}$_{diff}$. 
We used the Normalized Root Mean Squared Error (NRMSE) across dictionary entries to evaluate the accuracy of the result outputs.

\subsection{In vivo validation study} \label{MRF}
We evaluated the accuracy of our simulation method against the brute-force method in an MR Fingerprinting\cite{ma_magnetic_2013} application, in healthy rats. The sets of MR signals computed for the in silico validation of our approach are used here as MRF dictionaries. 

\subsubsection{Animal models} 
% infos du papier: Mapping of brain tissue hematocrit in glioma and acute stroke using a dual autoradiography approach avec B.Lemasson
Wistar rats (N=8, 268$\pm$23g) were imaged with the below-described MR protocol. All procedures were reviewed and approved by the local ethics committee (Comité éthique du GIN n°004), were performed under permits 380820 and A3851610008 (for experimental and animal care facilities) from the French Ministry of Agriculture (Articles R214–117 to R214–127 published on February 7, 2013), and reported in compliance with the ARRIVE guidelines (Animal Research: Reporting in Vivo Experiments).

\subsubsection{MRI acquisition}
MRI was conducted with a horizontal bore 4.7 T Biospec animal imager (Bruker Biospin, Ettlingen, Germany; IRMaGe facility) with an actively decoupled cross-coil setup (body coil for radiofrequency transmission and quadrature surface coil for signal reception) and Paravision 5.0.1. 

A GESFIDSE sequence (repetition time TR=4000 ms, 32 gradient-echoes, $\Delta$TE=3.3ms, Spin-Echo=60ms; NEX=1, 5 slices, FOV=30×30 mm$^2$, matrix=128×96 zero-filled to 128x128 and voxel size=234×234×800$\mu m^3$, acquisition duration 6min 24sec) was performed before and after the manual injection of the ultrasmall superparamagnetic iron oxide nanoparticles (USPIO) P904 (200$\mu$mol/kg body weight; Guerbet, Roissy, France). A three-minute delay after the first injection was applied before starting the second GEFIDSE acquisition.

Anatomical T2-weighted (T2w) images were acquired for ROI delineation, using a turbo spin-echo MRI sequence (TR=4000ms, echo-time TE=33ms, NEX=2, 31 slices, FOV=30×30mm$^2$, matrix=128×128 and voxel size=234×234×800$\mu m^3$, acquisition duration 4min 17sec).

\subsubsection{Image analysis}
Quantitative parameter maps of CBV, R, SO$_2$ and T2 were computed from the GESFIDSE data acquired on healthy animals, relying on a standard dictionary-matching process (inner-product, \cite{ma_magnetic_2013}) made on MP3\cite{brossard_mp3_2020}, a Matlab-based image processing software, using an in-house fingerprinting extension module.

In the three first validation studies, MRF parameter maps were reconstructed 
\begin{enumerate}
    \item with a 20,000-entry MRF dictionary simulated without diffusion effect using brute-force method (\textit{Ref}$_{nodiff}$) versus MR-WAVES (\textit{MRWAVES}$_{nodiff}$)
    \item with a 2,000-entry MRF dictionary with diffusion computed using the brute-force method (\textit{Ref}$_{diff}$) versus the proposed GRU-RNN for diffusion only (\textit{GRU}$_{diff}$) 
    \item with a 2,000-entry MRF dictionary with diffusion computed using the brute-force method (\textit{Ref}$_{diff}$) versus the whole MR-WAVES (\textit{MRWAVES}$_{diff}$)
\end{enumerate} 
Results using a dictionary computed with the proposed and the brute-force methods were compared using relative difference and Bland-Altman analysis.  

Finally, as the speed of MR-WAVES simulations allows large-scale dictionary generation, we used it to produce an MRF dictionary of 135,000 MR signals computed using 45,000 voxels with varying [35-95\%] SO$_2$ and [45-150ms] \tdeux{} values. Mean parameter values were computed in manually drawn ROIs: skull-extracted brain, cortex and striatum areas. Due to the long computation time of the brute-force method, no reference dictionary of 135,000 MR signals was computed.

%%%%%%%%%%%%%%%%%%%%%%%%%%%%%%%%%%%%%%%%%%%%%%%%%%%%%%%%%%%%%%%%%%%%%
%%%%%%%%%%%%%%%%%%%%%%%%%%%%%%%%%%%%%%%%%%%%%%%%%%%%%%%%%%%%%%%%%%%%%
%%%%%%%%%%%%%%%%%%%%%%%%%%%%%%%%%%%%%%%%%%%%%%%%%%%%%%%%%%%%%%%%%%%%%

\section{Results}

\subsection{Brute force 3D simulations}\label{result_refnodiff}
Excluding the computation of the 3D magnetic field inhomogeneities matrices, the brute force 3D simulations \textit{Ref}$_{nodiff}$ and \textit{Ref}$_{diff}$ compute respectively at a speed of 3 and 0.2 signals/seconds (\textit{i.e.} 6,700 and 90,200 seconds for 20,000-entry datasets). For the reference method, the computation time increased linearly with the number of signals simulated (data not shown).

\subsection{Fast 1D simulations without diffusion effect}
\fig{Figure 2} highlights the effect of the microvascular characteristics of the voxel on the 1D frequency histogram. Changes in the CBV and R values (here related to data-augmented voxels) have a strong effect on the frequency distributions. The sensitivity to this effect is of high importance to maintain the accuracy of the simulation compared to fully resolved 3D matrices.

After conducting numerous experiments adjusting the sampling of the 3D matrices using 1D distribution histograms, the optimal strategy that balances computational efficiency and accuracy was identified. The results, shown in \fig{Figure 5}, highlight the final approach based on a specific number of bins and bin edges that effectively represent the optimal frequency "step" in Hz, which is the focus of our analysis. Only the most relevant tests are presented, computed for fixed histogram edges (-200;200Hz) and the adapted number of bins to vary the sampling step. 

As it could be expected, \fig{Figure 5} shows that the NRMSE between the brute-force method (\textit{Ref}$_{nodiff}$) and our proposed approach (\textit{MRWAVES}$_{nodiff}$) decreases with smaller sampling step sizes, reflecting a more accurate description of the 3D matrices using less discretized 1D histograms.

As described previously (\ref{microvasc}), the magnetic susceptibility of the vascular component is adjusted to account for the contrast-agent effect, which broadens the frequency distribution, as can be seen in \fig{Figure 1}. Given that the Full-Width at Half Maximum (FWHM) of the post-CA distributions is larger than that of the pre-CA distributions, a larger frequency interval (\textit{i.e.} histogram edges) is necessary for post-CA signals to ensure a realistic description of the voxel’s frequency inhomogeneity (see \fig{Figure 5} where NRMSE stabilizes at larger step sizes for post-CA signals).

Based on these observations, we set the 1D distribution parameters as follows: for pre-CA, a range from -100 Hz to 100 Hz with a 0.5 Hz step; for post-CA, a range from -200 Hz to 200 Hz with a 1 Hz step. This approach balances computational efficiency and simulation accuracy, yielding an NRMSE of 0.07\% for pre-CA and 0.16\% for post-CA compared to the fully resolved 3D matrix method. On the signal concatenating pre and post-CA parts, the normalization slightly increased the NRMSE to 0.26\%. 

%A more detailed analysis of the error across the parameter space is provided in \fig{Figure 5b}. For post-CA, the error is well-distributed overall parameter values. For pre-CA, it could be noticed that higher CBV values lead to higher NRMSE, related to the broadening of the distributions for high CBV values which overflow the edges of the histograms. 

Using this description, the $MRWAVES_{nodiff}$ method computes 20,000 MR signals with varying microvascular properties (SO$_2$, CBV, R) and T2 in 32 seconds. This corresponds to a simulation speed of 625 signals per second, achieving a 200-fold acceleration compared to the \textit{Ref}$_{nodiff}$ approach (\ref{result_refnodiff}).

Using the GESFIDSE data acquired on the 8 animals of our study, MRF parameters maps of (CBV, R, SO$_2$ and \tdeux{}) are reconstructed in 20 seconds of matching time per animal, using dictionaries computed with both methods. In \fig{Figure S1}, examples of reconstructed MRF maps are shown for one animal of our study, comparing the reconstruction made with the brute force and our proposed method using frequency histogram distributions. Relative difference maps are shown on the bottom line. Density histograms of the relative difference across the voxels of the slices are also shown, exhibiting a strong similarity between both methods.

\subsection{GRU-RNN simulations of the water diffusion effect on MR signals}

The training on the GRU RNN lasts approximately 300 seconds. On the 2,000 testing signals, the prediction time is just 3 seconds. Considering that input signals are here computed using \textit{Ref}$_{nodiff}$, it means a total simulation time of 6,703 seconds, meaning a speed of 3 signals per second compared to 0,2 signals per second for the \textit{Ref}$_{diff}$ brute force method. \fig{Figure 6a} illustrates examples of predicted signals, comparing them with both the input signals (without diffusion) and the expected ground truth for the pre and post-CA sections. It can be seen that the prediction closely matched the ground truth signals from \textit{Ref}$_{diff}$. 
\fig{Figure 6b} presents the normalized root mean square error (NRMSE) calculated between the 2,000 predicted signals (\textit{GRU}$_{diff}$) and their corresponding ground truth (\textit{Ref}$_{diff}$) across the parameter space. The overall error is low (<4\%) and uniformly distributed across different parameter values. However, notable exceptions are observed for signals corresponding to large vessels (R>$15\mu$m and CBV>5\%), where the error is highest. These cases can be considered outliers, as such values are uncommon in healthy rat brains.

In \fig{Figure S2}, we show an example of the results on one slice from one of the 8 healthy animals. The relative difference maps highlight the similarity between the results using both simulation methods.

\subsection{Complete MR-WAVES method}

\subsubsection{\textit{MRWAVES}$_{diff}$ vs \textit{Ref}$_{diff}$ for small MRF dictionary generation}
The mean NRMSE between 2,000 signals (pre and post-CA concatenation) simulated with the brute-force method (\textit{Ref}$_{diff}$) on one side and MR-WAVES (\textit{MRWAVES}$_{diff}$, combining 1D histograms and GRU-RNN) on the other side is 0.26\%, reflecting that the main error provides from the GRU-RNN for which error was also near 0.26\%. Using MR-WAVES, simulation time is less than 5 seconds including fast microvascular simulations and GRU-RNN inference time, \textit{i.e.} a simulation speed of around 400 signals per second. For a small dictionary, loading the external data and the inference model and saving the signals used a significant part of the simulation time, artificially decreasing the simulation speed.

In \fig{Figure 7a}, MRF maps of CBV, R, SO$_2$ and \tdeux{} computed with a dictionary with diffusion from the \textit{Ref}$_{diff}$ method and the proposed MR-WAVES dictionary combining fast simulations and fast diffusion representation (\textit{MRWAVES}$_{diff}$) are shown for single-slice examples of one animal, supplemented with relative difference maps.

Visually, the resulting maps computed from the MR-WAVES model are of the same quality as the reference maps. This observation is confirmed by Bland-Atlman plots in the whole brain for all animals in \fig{Figure 7b}. For the CBV, the plot shows a small bias of around -0.15, indicating that MR-WAVES tends to slightly overestimate CBV compared to the brute-force method. The limits of agreement ($\pm$1.96 SD) range is narrow, suggesting a good level of agreement between the methods. For the R estimates, the healthy population plot reveals a larger bias of approximately -0.4, with all data points within the limits of agreement, though some variance is observed. For the SO$_2$ parameter, analysis shows a minimal bias and excellent agreement between methods, as the differences hover around zero. For \tdeux{}, the healthy population shows good agreement between methods, with a bias close to zero and minimal spread, indicating consistency. 
Overall, the Bland-Altman analysis highlights good agreement between the brute-force method and MR-WAVES.

\subsubsection{\textit{MRWAVES}$_{diff}$ for large MRF dictionary generation}

Finally, a large MRF dictionary of 135,000 entries is studied. Using MR-WAVES and considering that intra-voxel frequency distributions are precomputed for a large database of tissue parameters, the generation of the dictionary took 74 seconds for pre and post-CA signals including the inference of the RNN in 22 seconds. Based on our observation of the simulation time using the brute-force method linearly increasing with respect to the number of simulated signals, the equivalent simulation of an MRF dictionary of 135,000 entries would take 278 hours. Thus, we observe an acceleration factor of 13,000 with the proposed MR-WAVES method. 

From this simulated large dictionary, MRF maps of CBV, R, SO$_2$ and \tdeux{} on animal models are computed and shown for one slice of each animal in \fig{Figure 8a}.

It can be observed that the results are different from those obtained with smaller dictionaries (\fig{Figure 7a}), especially for estimated values of R and SO$_2$ that seem to be increased whereas the CBV and \tdeux{} values are consistent with previous observations but smoother because the sampling of the parameters in the dictionary is thinner. In \fig{Figure 8b}, detailed estimation values are plotted for all animals with the corresponding standard deviation. A detailed description of individual parameter estimated values is provided in \fig{Table 1}. MRF estimates of microvascular and relaxometry parameters are consistent across animals and in line with expected literature values \cite{delphin_enhancing_2024}.

%%%%%%%%%%%%%%%%%%%%%%%%%%%%%%%%%%%%%%%%%%%%%%%%%%%%%%%%%%%%%%%%%%%%%
%%%%%%%%%%%%%%%%%%%%%%%%%%%%%%%%%%%%%%%%%%%%%%%%%%%%%%%%%%%%%%%%%%%%%
%%%%%%%%%%%%%%%%%%%%%%%%%%%%%%%%%%%%%%%%%%%%%%%%%%%%%%%%%%%%%%%%%%%%%

\section{Discussion and Conclusions}

In this paper, we introduced MR-WAVES, a novel simulation method for incorporating water-diffusion and vascular effects into MR signal simulations. By combining a simplified representation of intra-voxel magnetic field inhomogeneities with the GRU-RNN, MR-WAVES achieves over 13,000 times the computational speed of previously proposed brute-force methods while maintaining the accuracy of the signal simulations. 

In silico studies show that concerning the first fast simulation tools utilizing pre-computed intra-voxel frequency distributions, our results align closely with brute-force simulation outputs, as confirmed by statistical analyses on simulated MR signals.

For the RNN implementation, two models were trained on separate datasets for pre-CA and post-CA signals, following the two-stage acquisition protocol. Since the recurrence in the RNN model takes into account time-dependent data points in the processed inputs, this approach reduces errors by decoupling the last time point of the pre-CA signal from the first time point of the post-CA signal. The predicted outputs showed an NRMSE of up to 0.26\% compared to deterministic simulations of water diffusion. 
 
The MRF in vivo validation study that we conducted used the proposed simulation methods for the computation of parametric maps for CBV, R, SO$_2$, and \tdeux{} while accounting for the effects of water diffusion.
With the proposed GRU-RNN, the reconstruction was effective in generating accurate quantitative maps for CBV, R, SO$_2$, and \tdeux{}. The slight prediction errors appear to have a limited impact on matching, especially given the noise in acquired signals due to undersampling. However, during RNN training, we observed minor overfitting, which could contribute to prediction errors for data outside the training parameter space. To address this, we propose to further employ data augmentation techniques or add dropout layers to the RNN architecture. Additionally, for longer sequences like MRF-bSSFP, More robust networks like LSTMs \cite{barrier_marvel_2024} may be considered for longer sequences to improve the diffusion effect simulations.

When the full protocol MR-WAVES was tested against a brute-force method with diffusion on a dictionary of 2,000 entries, differences in the reconstructed parametric maps were a little bit more pronounced, as expected. The approximation of intra-voxel inhomogeneities in the first part of MR-WAVES could contribute to error propagation through the RNN. 
Because relaxation was modeled using a simple exponential form, introducing bias from prior simulation errors, we believe that incorporating \tdeux{} from the beginning of the Bloch simulations could further reduce the estimate errors, although this would extend simulation time. 

In the in vivo MRF application involving a larger dictionary, R and SO$_2$ values were higher than those observed in the 2,000-entry experiment. This could result from differences in parameter distributions. With the 135,000-entry dictionary, our results are in line with the experiments conducted by \cite{delphin_enhancing_2024} that have validated their method against histological and analytical estimations. Our SO$_2$ estimates specifically align in cases with a simplified cylindrical vascular model, though they exceed those derived from more realistic microscopy-based structures. This suggests that our non-spatial distribution function for field inhomogeneities might oversimplify the microvasculature representation and it supports the idea of increasing the density of the offset frequencies simulated before convolution with the frequency histogram.

Future work will focus on adapting MR-WAVES to incorporate multiple ADCs in the MR simulations framework. Preliminary tests suggest that certain neural network architectures can effectively learn water diffusion representations across varying ADC values. This extension has the potential to improve microvascular measurements in pathological tissues such as stroke or tumor, where the fixed ADC value of 850 $\mu$m$^2$/s may no longer be appropriate \cite{sener_diffusion_2001} within the MRF dictionary. This could also enable the simultaneous estimation of ADC maps as an additional MRF parameter, which is particularly relevant for stroke diagnosis. 

Finally, our proposed MR-WAVES method appears to be generalizable to a broad range of MR sequences. Notably, in a previous study\cite{coudert_relaxometry_2025}, fast microvascular simulations based on 1D frequency distributions were successfully implemented with balanced steady-state free precession (bSSFP) sequences. This approach enabled non-contrast perfusion estimates in human volunteers, leveraging simplified frequency distributions for microvascular modeling. Our initial observations suggest that the water-diffusion effect model can be rapidly trained on signals from sequences other than GESFIDSE and produces similar results accuracy. This also opens the possibility of developing more robust neural network architectures capable of learning signal representations from multiple sequences simultaneously. Such an approach could provide a generalizable simulation tool however leading to longer training times and possibly reduced estimation accuracy.

The proposed MR-WAVES also aligns with the ongoing efforts in microstructure MRI to refine model-based tissue characterization \cite{zhang_noddi_2012, assaf_axcaliber_2008, assaf_composite_2005, novikov_present_2021}. By efficiently simulating diffusion processes at different scales, it could contribute to resolving key challenges outlined in recent literature, such as the identification of intra- vs. extra-cellular diffusion components and the MRI characterization of microstructures.

In summary, the proposed tool provides a powerful framework for fast MR simulations in microvascular structures considering the water-diffusion effect. Fast numerical computations of such tissue properties effects serve to improve both our understanding of the BOLD effect and the quantification of microvascular tissue parameters using MRF methods. By enabling more flexible and computationally efficient simulations, it has the potential to broaden the application of MRF-based techniques and microstructure diffusion-MRI.

%%%%%%%%%%%%%%%%%%%%%%%%%%%%%%%%%%%%%%%%%%%%%%%%%%%%%%%%%%%%%%%%%%%%%
%%%%%%%%%%%%%%%%%%%%%%%%%%%%%%%%%%%%%%%%%%%%%%%%%%%%%%%%%%%%%%%%%%%%%
%%%%%%%%%%%%%%%%%%%%%%%%%%%%%%%%%%%%%%%%%%%%%%%%%%%%%%%%%%%%%%%%%%%%%
\subsection*{Author contributions}
All authors listed have made substantial, direct, and intellectual contributions, proofread and corrected the final manuscript, and approved it for publication.

\subsection*{Financial disclosure}

The MRI facility IRMaGe is partly funded by the French program “Investissement d’avenir” run by the French National Research Agency, grant “Infrastructure d’avenir en Biologie et Santé”. [ANR-11-INBS-0006]
The project is supported by the French National Research Agency. [ANR-20-CE19-0030 MRFUSE]

\newpage
% \bibliography{biblio}%
\vfill\pagebreak

\section*{Supporting information}
The following supporting information is available as part of the online article:

\vskip\baselineskip\noindent
\textbf{Figure S1.} shows reconstructed MRvF maps computed using a 20,000-entry zero-diffusion dictionary from the reference and for the proposed methods. For this one animal, results are compared through a one-slice relative difference map and an all-slices relative difference histogram for each estimated parameter. Bland-Altman analyses across all animals of the study are also shown in the whole brain ROI.

\noindent
\textbf{Figure S2.} shows MRvF parametric maps reconstructed from a 2000-entry dictionary computed with zero-diffusion using brute force method (top), with an ADC of 850$\mu m/s^2$ using brute force method i.e. Gaussian kernel (middle) and with an ADC of 850$\mu m/s^2$ using the proposed GRU-RNN (bottom). Relative difference maps are also shown for each estimated parameter.

\vspace*{6pt}

\newpage

\listoffigures
\listoftables

\begin{figure*}
\centerline{\includegraphics[  width=\textwidth  ]{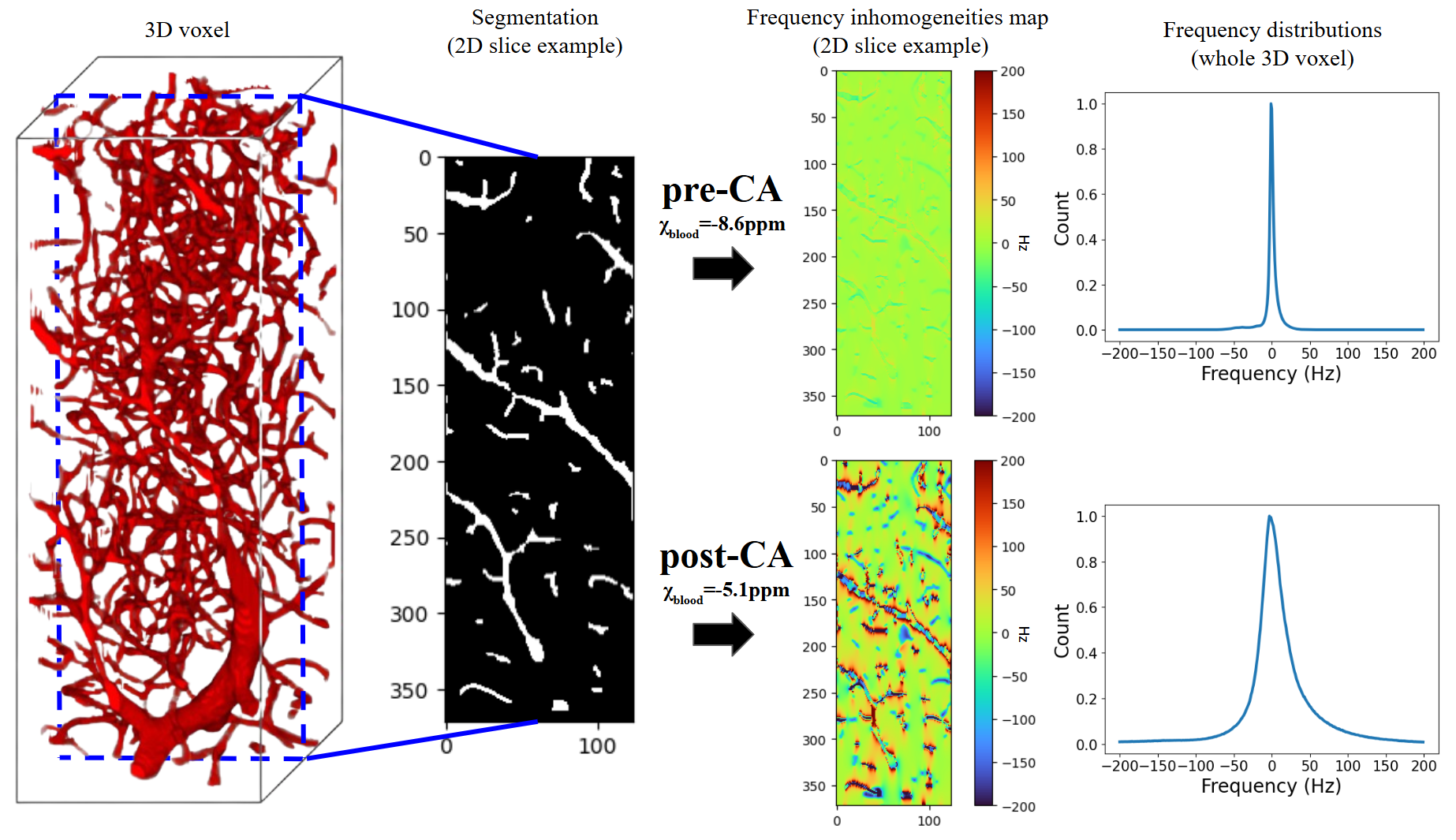}}
\caption{ Illustrated example of frequency inhomogeneities distribution computation. A microscopy voxel is binary segmented, and susceptibility values associated with the vascular and extravascular components are then used to compute magnetic field inhomogeneity maps. The equivalent frequency inhomogeneity maps are saved as 3D frequency distribution in the voxel.\label{fig1}}
\end{figure*}

\begin{figure*}
\centerline{\includegraphics[  width=\textwidth  ]{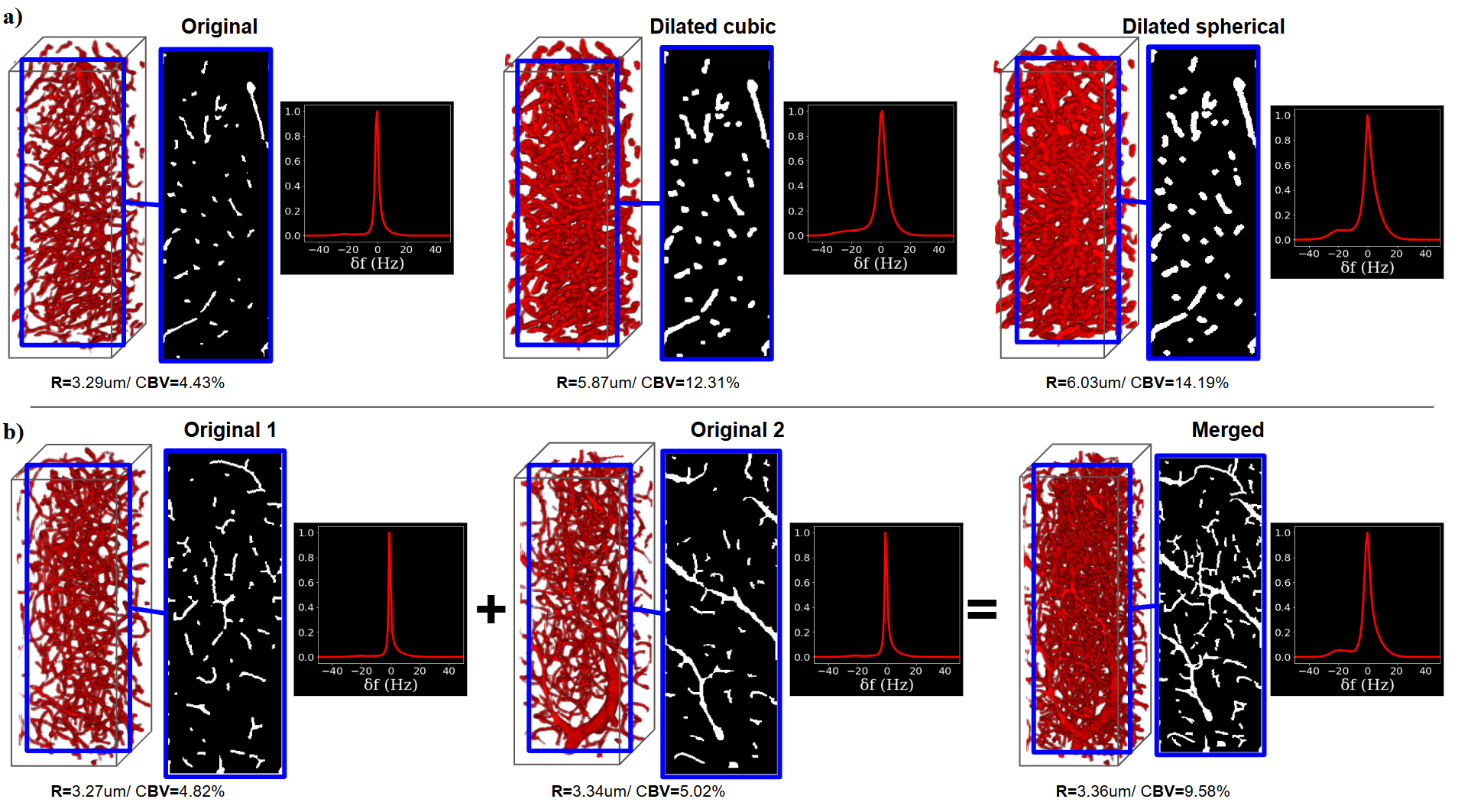}}
\caption{ Microvascular voxel data augmentation. Examples of augmentation processes, a) dilation with a cubic kernel (top middle), dilation with a spherical kernel (top right)  and b) merging of two voxels (bottom line). For each case, a 3D view of the microstructure, a 2D segmentation example, and a 1D frequency distribution are shown.\label{fig2}}
\end{figure*}

\begin{figure*}
\centerline{\includegraphics[  width=\textwidth  ]{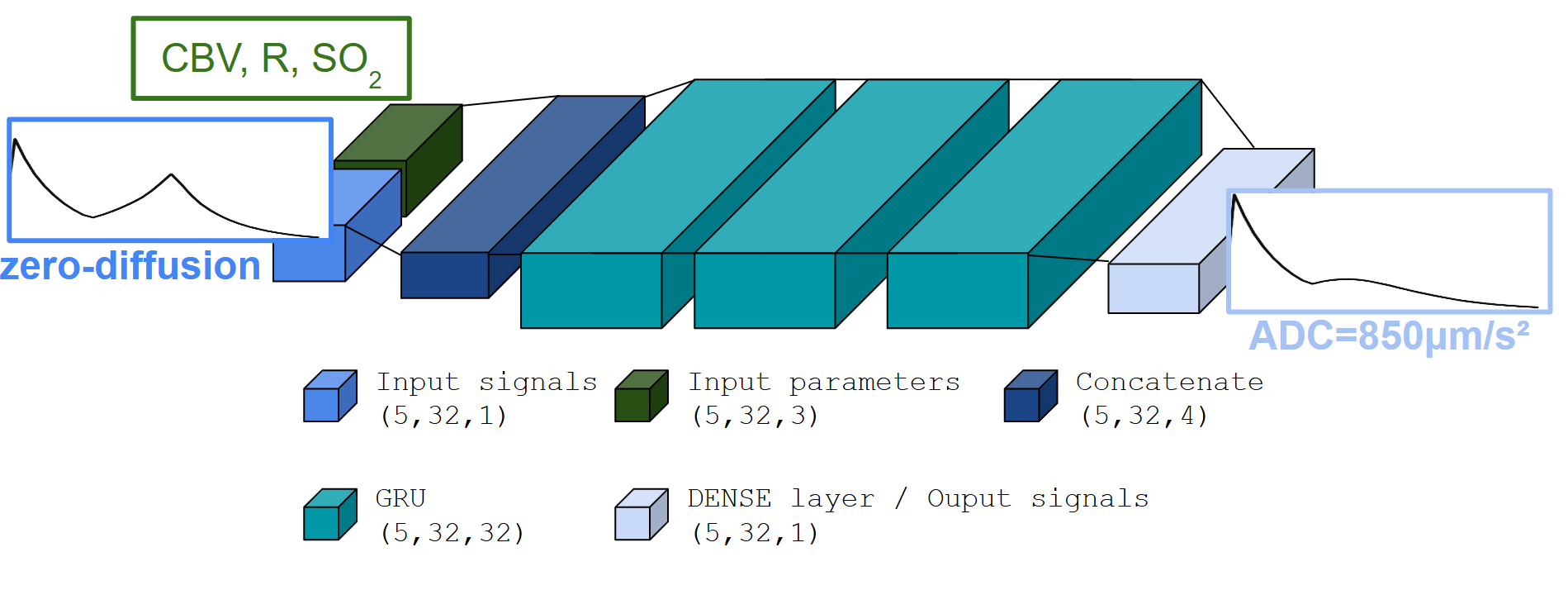}}
\caption{Schematic illustration of the GRU-RNN layers architecture. The zero-diffusion MR signal (pre or post-CA) of 32-time points is taken as input and concatenated with a triplet of microvascular parameters. The effect of water-diffusion for an AD of 850$\mu m / s^2$ is then applied through GRU layers leading to an output MR signals of 32 time points. \label{fig3}}
\end{figure*}

\begin{figure*}
\centerline{\includegraphics[  width=\textwidth  ]{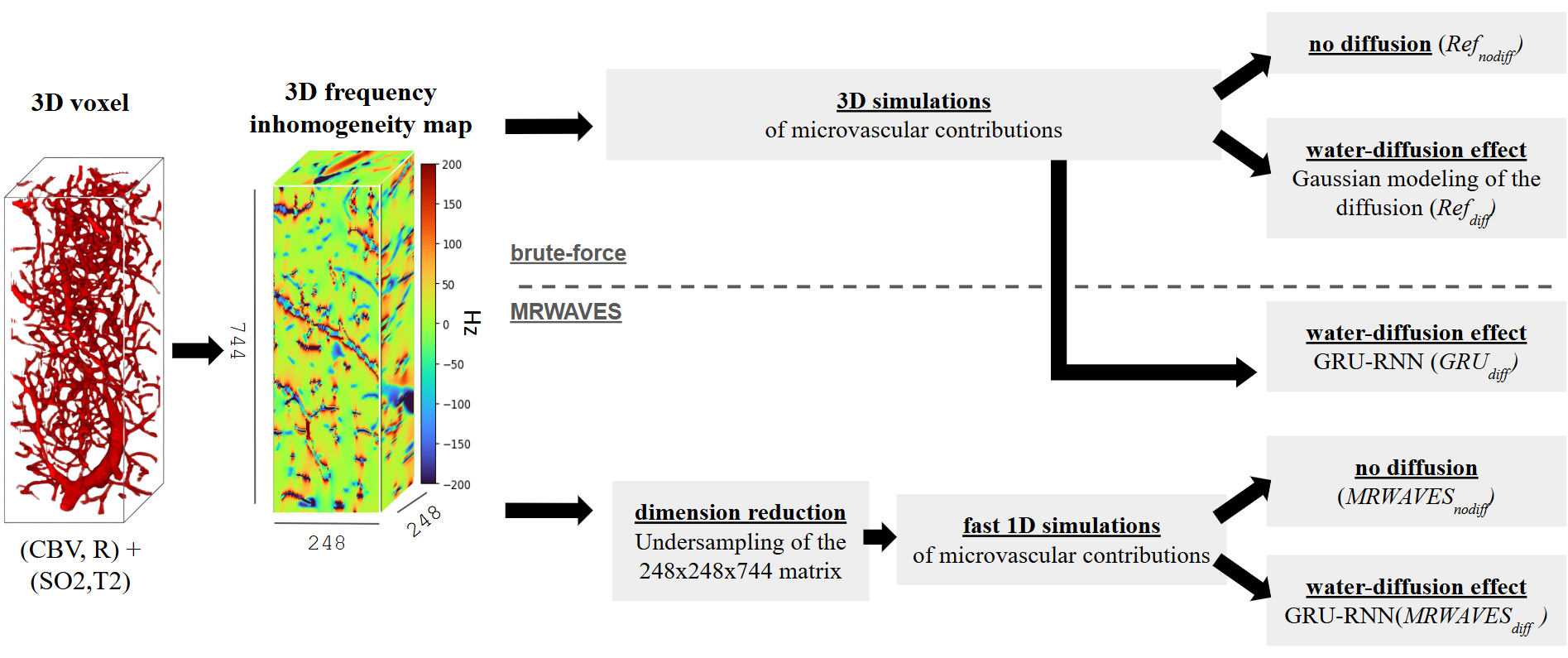}}
\caption{Schematic illustration of the proposed method compared to the reference method. The reference method relies on computed 3D frequency inhomogeneity maps with 248x248x744 intra-voxel contribution, to compute signals with a 0.5ms time-step to account for water-diffusion Gaussian motion. The proposed method undersamples the 3D inhomogeneity map to simplify computation and perform microvascular simulation at a time-step equal to the repetition time (TR), and the water-diffusion is a posteriori modeled by the RNN.\label{fig4}}
\end{figure*}

\begin{figure*}
\centerline{\includegraphics[  width=\textwidth  ]{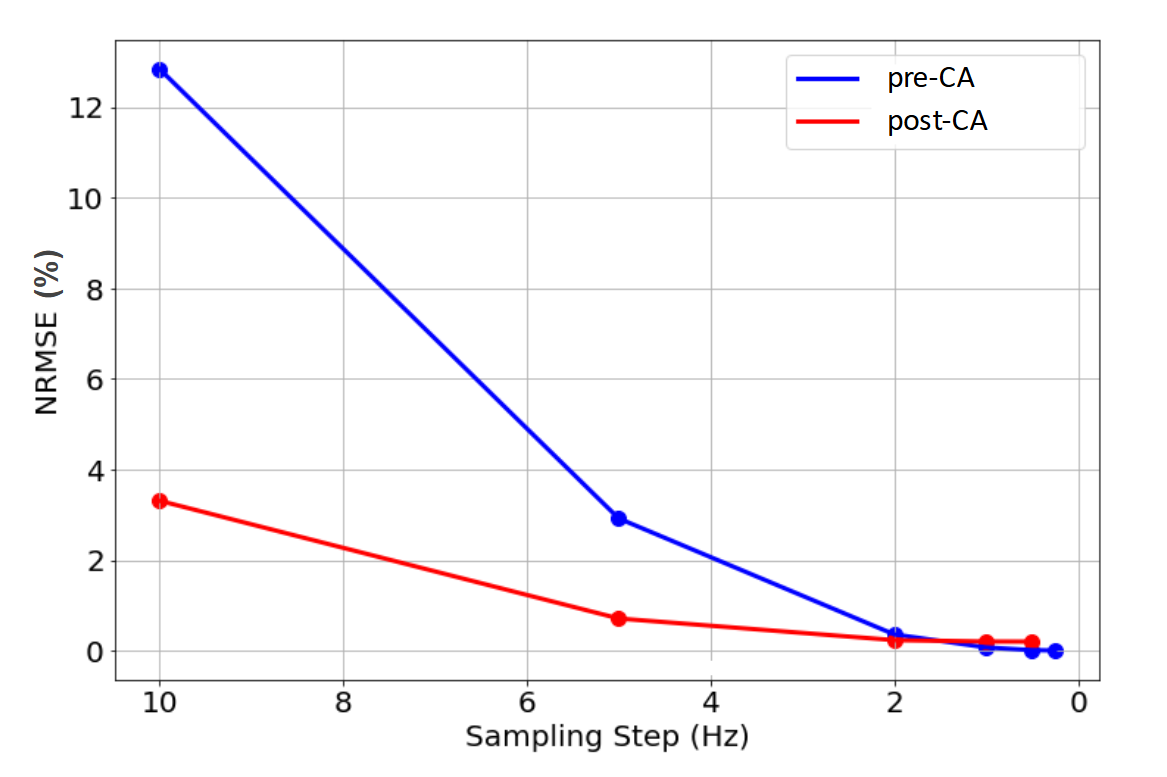}}
\caption{NRMSE computed between proposed ($MRWAVES_{nodiff}$ with several sampling) and reference ($Ref_{nodiff}$) dictionaries for pre (blue) and post-CA (red) simulated MR signals. NRMSE is computed for different sampling steps (from 10Hz to 0.5Hz for post-CA and 0.25Hz for pre-CA) of the histogram frequency distribution into a microvascular voxel. As the the number of bins in the histogram increases, the sampling step decreases and so does the NRMSE between the two methods.\label{fig5}}
\end{figure*}

\begin{figure*}
\centerline{\includegraphics[  width=\textwidth  ]{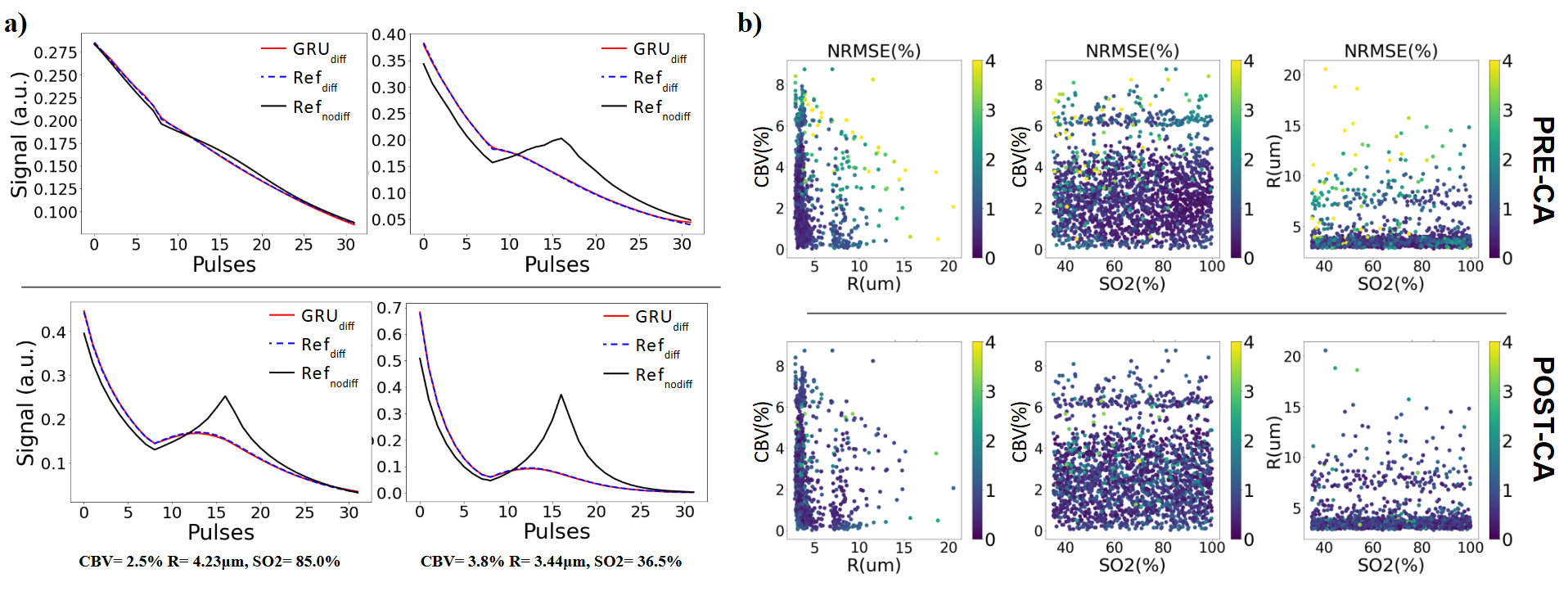}}
\caption{2,000 signals without diffusion computed with the brute force method (from $Ref_{nodiff}$) are used as input in the GRU-RNN. Results are compared with brute force signals computed with diffusion (from $Ref_{diff}$). a) GRU-RNN prediction examples ($GRU_{diff}$) for two voxels with fixed CBV, R and SO2 values where the signal is computed without and with CA contribution. Ground truth signal as well as zero-diffusion signal used for the prediction are shown. b) NRMSE between GRUdiff prediction and Refdiff ground truth, across the whole parameter space. \label{fig6}}
\end{figure*}

\begin{figure*}
\centerline{\includegraphics[  width=\textwidth  ]{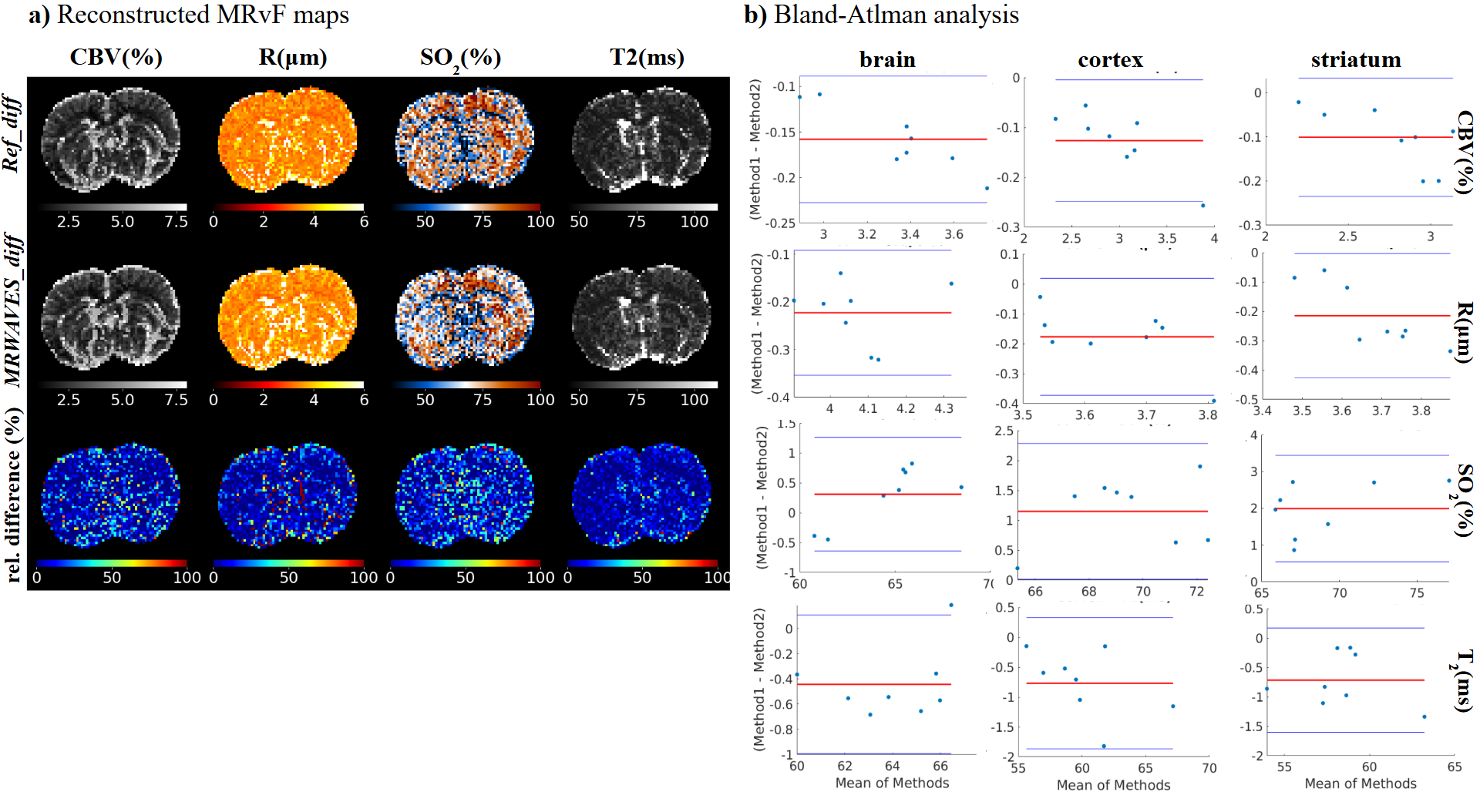}}
\caption{ a) One slice example of MRvF parametric maps reconstructed from a 2,000-entry dictionary computed for an ADC of 850$\mu m / s^2$ using the brute force method ($Ref_{diff}$, top) and the proposed MRWAVES ($MRWAVES_{diff}$, bottom). Relative difference maps are also shown for each estimated parameter. b) Bland-Altman analyses across all animals of the study in three ROIs of the brain. Bias is indicated by the red line. Limits of agreements (±1.96 SD) are indicated by the blue lines. Each point represents one animal.
\label{fig7}}
\end{figure*}

\begin{figure*}
\centerline{\includegraphics[  width=\textwidth  ]{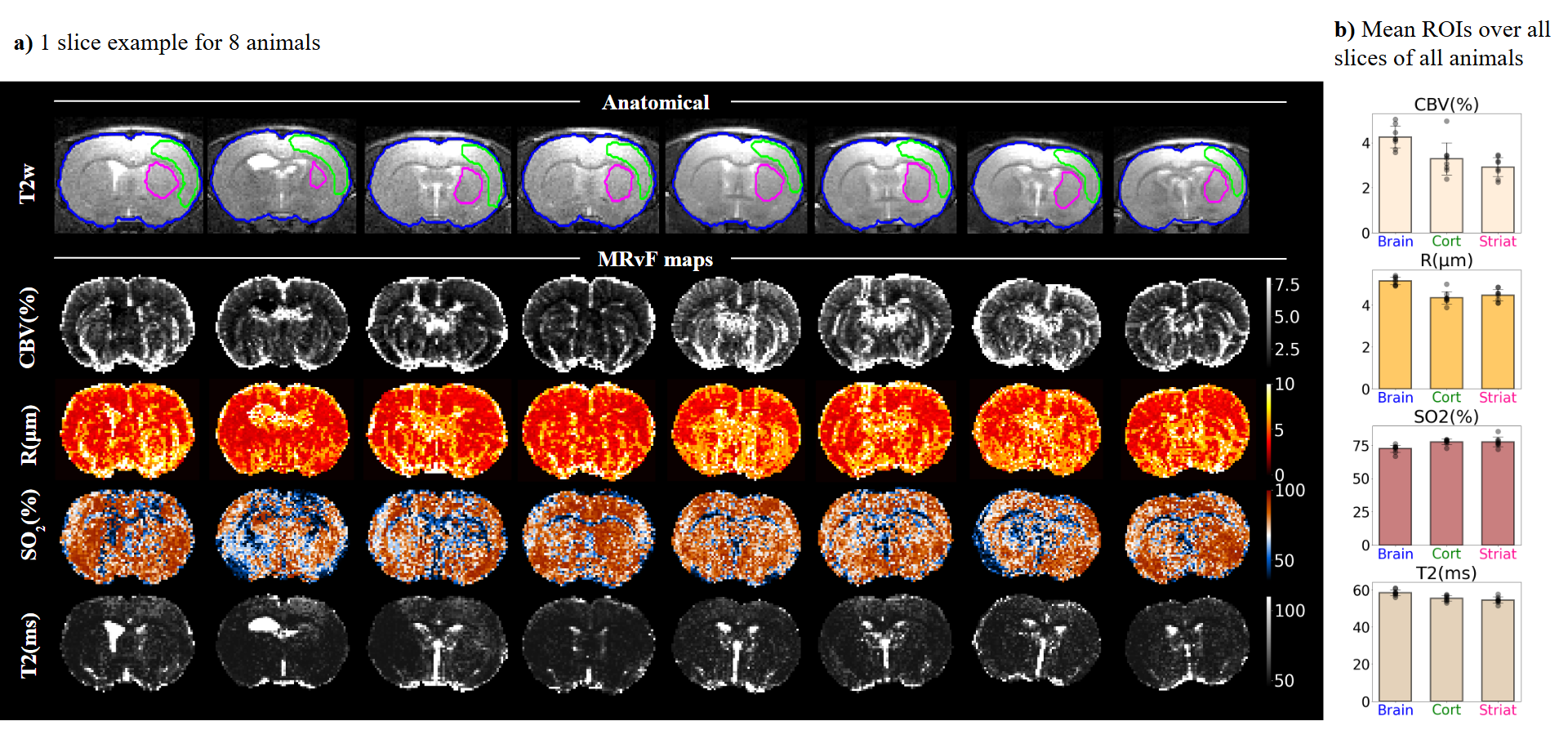}}
\caption{ a) One-slice MRvF parametric maps for each animal of the study, reconstructed using a 135,000-entry dictionary computed with $MRWAVES_{diff}$. b) Mean and standard deviation of parameter estimates in brain ROIs displayed on the anatomical T2w scan on the left. \label{fig8}}
\end{figure*}

\begin{table*}[h!]
    \centering
    \caption{ROI Values for whole Brain, Cortex, and Striatum for each animal of the study (computed with the 135,000 entry MR-WAVES dictionary).}
    \begin{tabular}{lcccc}
        \toprule
        Region & CBV (\%) & R ($\mu$m) & SO$_2$ (\%) & T$_2$ (ms) \\
        \midrule
        Brain & 4.1590 & 5.2948 & 73.2511 & 60.6642 \\
                    & 4.8557 & 5.4027 & 66.7667 & 61.1601 \\
                    & 3.7131 & 4.8959 & 72.2098 & 57.9048 \\
                    & 3.5615 & 4.9131 & 76.0095 & 56.0804 \\
                    & 4.4469 & 5.1905 & 73.5398 & 58.3167 \\
                    & 4.0613 & 4.9498 & 73.9492 & 57.1638 \\
                    & 5.0453 & 5.3207 & 69.9437 & 58.7428 \\
                    & 4.1632 & 5.1035 & 73.4670 & 58.2198 \\
        \midrule
        Cortex      & 3.3207 & 4.5049 & 77.8120 & 57.2103 \\
                    & 2.8993 & 4.3204 & 72.7278 & 56.5981 \\
                    & 2.7885 & 4.2270 & 79.3297 & 53.9291 \\
                    & 2.3842 & 3.8678 & 78.8777 & 53.0398 \\
                    & 3.4252 & 4.2866 & 78.2382 & 54.5659 \\
                    & 3.4364 & 4.3209 & 77.6453 & 55.9036 \\
                    & 4.9560 & 4.9892 & 76.4490 & 57.5645 \\
                    & 3.0758 & 4.2030 & 78.9589 & 54.7975 \\
        \midrule
        Striatum    & 2.8166 & 4.4798 & 72.0547 & 55.9499 \\
                    & 3.4512 & 4.8342 & 76.1648 & 54.3757 \\
                    & 2.3605 & 4.0911 & 75.2158 & 54.7879 \\
                    & 2.2498 & 4.0942 & 85.2227 & 51.4786 \\
                    & 3.1726 & 4.5476 & 77.0392 & 53.3658 \\
                    & 3.1312 & 4.5599 & 78.3934 & 54.1147 \\
                    & 3.4182 & 4.8302 & 76.3880 & 57.6723 \\
                    & 2.7379 & 4.2582 & 79.0244 & 54.2772 \\
        \bottomrule
    \end{tabular}
\end{table*}

\FloatBarrier
\bibliographystyle{plain}
\bibliography{biblio}

\includepdf[pages=-]{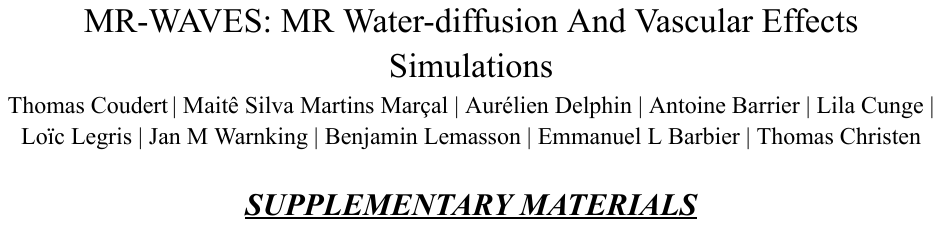} %
\end{document}